# A MODEL OF OPENEHR BASED ELECTRONIC MEDICAL RECORD IN INDONESIA

[1]A.B. Mutiara, [2]A. Muslim, [3] T. Oswari, [4] R.A. Miharja

[1,2,4]*Faculty of Computer Science and Information Technology, Gunadarma University, Indonesia*
[3]*Faculty of Economics, Gunadarma University, Indonesia*
E-mail:  [1,2,3]{amutiara,amuslim,toswari@staff.gunadarma.ac.id}, [4]iyand_auror@yahoo.com

**Abstract**

For the realization of the vision and mission of Healthy Indonesia 2015, we need a health service with a broad and comprehensive scope.To provide health services, it can be realized by creating an integrated information system applications such as creating an electronic medical record that has the ability to process and store patient medical data. The specifications used medical record is an open specification contained in OpenEHR that includes information and service model for electronic medical records, demographics, and the archetype which allows software developers taking the logical structure as a universal functional   interface, so it can facilitate the process of information by the recipient. It is because of using the interface with appropriate-purposed data presentation and data on computer screen of the same users. The purpose of this paper is to create an electronic website for the medical record by using OpenEHR specifications for easy accessing, processing and storing the medical records by the actors that play a role in the data processing of medical records. With this application it is expected to be useful for data processing and health information gathering, thus to improve the quality of services that will impact the improved performance of the hospital management.  The improved performance of the hospital management will become a supporter of the vision and mission Healthy Indonesia 2015.

**Keywords**: Electronic Medical Records (EMR), Health Information System (HIS), OpenEHR, HL7

## 1. Introduction

To provide health services with a broad and comprehensive scope, it is needed to build an integrated information system like Health Information System (HIS). Technology provided by HIS is very useful for improving the quality of care that can improve the performance of the hospital management, such as auditable and accountable documentations, integrated medical records so as to speed up the access and exchange of patient's medical records between hospitals, community health centers and clinics. Thus it can be achieved the vision and mission Healthy Indonesia 2015.

To realize the benefits of HIS technology can be implemented by creating an electronic medical record (Electronic Health Record) that has the ability to store patient medical data. Specifications of the proposed medical record are an open specification contained in OpenEHR. OpenEHR specification enables the exchange of information between different medical record formats, such as HL7 and CEN / ISO [2,3].

Based on the background above, the authors try to achieve the mission Healthy Indonesia 2015 with creating an HIS application namely application of web-based electronic medical record to speed up the access and exchange of patient medical records between hospitals, community health centers and clinics according to the specifications that provided by OpenEHR.

## 2. Method And Design

Medical records containing information about the evaluation of physical condition and the history of patient are very important in the planning and coordination of patient care for further evaluation and to ensure continuity of services provided. The completeness, accuracy and timeliness of the filling should be pursued in healthcare organizations because it is very important for the feasibility of actions, and referral services.

Electronic Health Record (EHR) is an activity to computerize health records and process the content associated with it [1]. At first the medical record in Indonesia still known as medical records and even now some hospitals in Indonesia are still using the same term. Medical Record is "*a set of facts about the life history of the patient and the illness, including illness, current and past treatments are written by health professionals in their efforts to provide health care to patients*" [1]. EHR system is not information that can be purchased and installed as word-processing package or payment information and laboratory systems that are directly linked to other information systems and tools that fit in a particular environment. EHR is an information system that has a broader framework and meets a set of functions [3].

### 2.1 Waterfall Method

To create the application website electronic medical record based on OpenEHR it is needed a method of research that constitutes the data collection procedures used to solve the problem at hand. The scientific method conducted in this research is by using development software engineering Waterfall methods [7]. This method is often used by the analyzer system in general. The essence of the Waterfall method is construction of a system are performed sequentially or in a linear fashion. So if step one has not done, it will not to be able to do steps 2, 3 and so on. Automatically, phase 3 will be able to do if the phases 1 and 2 have been performed.

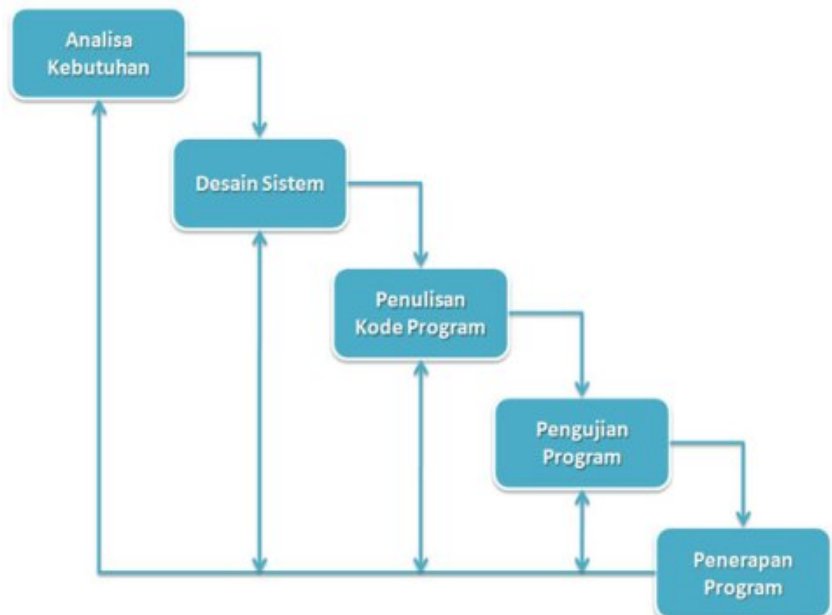

Figure 1: Steps of the Waterfall Method

Globally, the *Waterfall* method has undertaken steps are described as the following [7]:
1. **Analysis**

Analysis step is a needs analysis of the system. In this stage data is collected by giving the questionnaires to hospitals and polyclinics in Depok in order to know what the system has been running. To better understand about the data stream of medical records, then we did an interview with someone who had been studying the medical records. After that we did a literature study on the design of an electronic medical record system processing. This study was conducted to find out the specifications related OpenEHR how to classify medical records. Besides that literature study was also conducted to explore, learn and summarize a wide range of literature relating to the formulation of the problem, theories relating to the creation of electronic health records websites [5,6] such as electronic health records websites, *Yii Framework*[1,4,8], *WampServer*, Use-Case Diagram, Ontograf, Class Diagram and navigation structure. Stages of analysis would be a reference to translate into the language programmer.

2. **Design**

The design process can be used to translate the requirements needed for a software design that can be predicted prior to coding made. In this phase it is carried out the design of the new system that will be created. To simplify the representation system to be made, then it is designed a Use-Case Diagram, Ontograf, a Class Diagram and navigation structure. A Use-Case Diagram describes the expected functionality of a system. The emphasis is on "what" is done by the system, and not the "how". A Use-Case represents an interaction between the actors with the system. Ontograf is an ontology modeling to describe the conception of an integrated knowledge base. With Ontograf the sub-class relationship can be seen in the tree. It also supports a multiple inheritance, and root on the formed class hierarchy, namely the class "THING". While the Class Diagram describes the structure and description of classes, packages and objects and their relationships to each other such as containment, inheritance, associations and others. Then, we designed the navigation structure of the developed system. Navigation structure can be used to describe the used flow of created application. After all, it can be done designing the page of the built website. The used software architectures in building this website are: i) *Windows Seven Ultimate* Operation System; ii) *WampServer* 5 with *Apache* 2.2.4, *PHP* versi 5.2.2, and *MySQL* version 5.0.37; iii) *Yii Framework* version 1.1.10

3. **Coding and Testing**

Coding is the design translation into the programming language that can be recognized by the computer. This stage is a real stage in working out a system. It's mean that the use of computer will be maximized in this stage. The encoding is done using Yii Framework [4]. Once the coding is completed, the created system will be tested. Testing is done by using the web server WampServer. The program was tested on localhost first. The aim of testing is to find faults in the system and then it must be repaired.

4. **Application**

This stage can be said to be final stage in making a system. After doing the analysis, design and coding, the system will be used by the user. Before the user can use the system, the website that has previously been tested on localhost must be hosted on the real web-server.

5. **Maintenance**

Websites that are hosted and used in the future by users will definitely experience the change. These changes can occur because of an error or because the software must adapt to the new environment such as the presence of a new device or a new operating system or the user requires functional development. Therefore, the created website must be maintained on a regular basis.

## 3. EMR Website System

### 3.1 Uses Case Diagrams
*Uses case diagram* are used to describe the expected functionality of a system. A use case diagram represents the interaction between the actors with the system. Use-Case Diagram of the system could be seen in Fig.2.

There are five actors who could use this system, the administrators, patients, staff, physicians and the laboratory. This website uses login system, so the user must have a username and password before using the system. Administrators are in charge of making the users. The users get the username and password from the administrator. Especially for the patients, they must register in advance to the staff when they visit. Staff will provide a default username and password that will be changed by the patient. The website is built on the role capability (the ability of the role) of the user. The system only provides the facility in accordance with the users needed of the system, so not all users can use the entire facilities website.

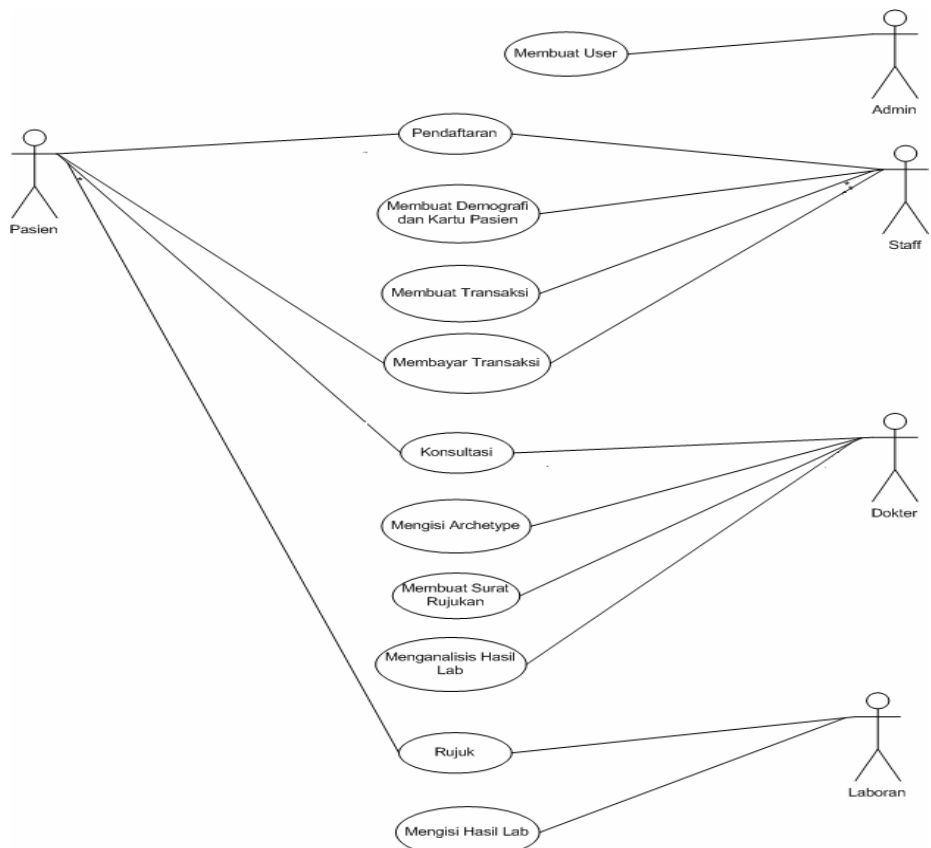

Figure 2: *Uses Case* Diagram of Website System

### 3.2 Class Diagram
*Class Diagram* is a diagram that describes the structure and description of classes, packages and objects as well as relationships with one another. Class Diagrams also explain the relationship between classes in a system and how they collaborate with each other in order to achieve a goal. Fig.3 shows the class diagram of the electronic medical record Web site.

From the Class Diagram it can be seen the relationship between the tables with other tables in the database. Patient table is a master table linked to a reference table. Reference table is a table that can not stand alone and is part of the master table. Reference table of the patients is a religion, insurance, sex, and status. The first patient has one religion, insurance, sex and status.

Patients have a relationship with the patient chart. It is found that 1 patient had a lot of patient cards. Patient card table is the master table. Patient and the patient's card are connected by Medical Record Number (MRN) which is a unique identity of the patient. The previous users will make the patient's card first. User is a master table which has a reference role. It is found that one user has one role. Users can create a patient chart.

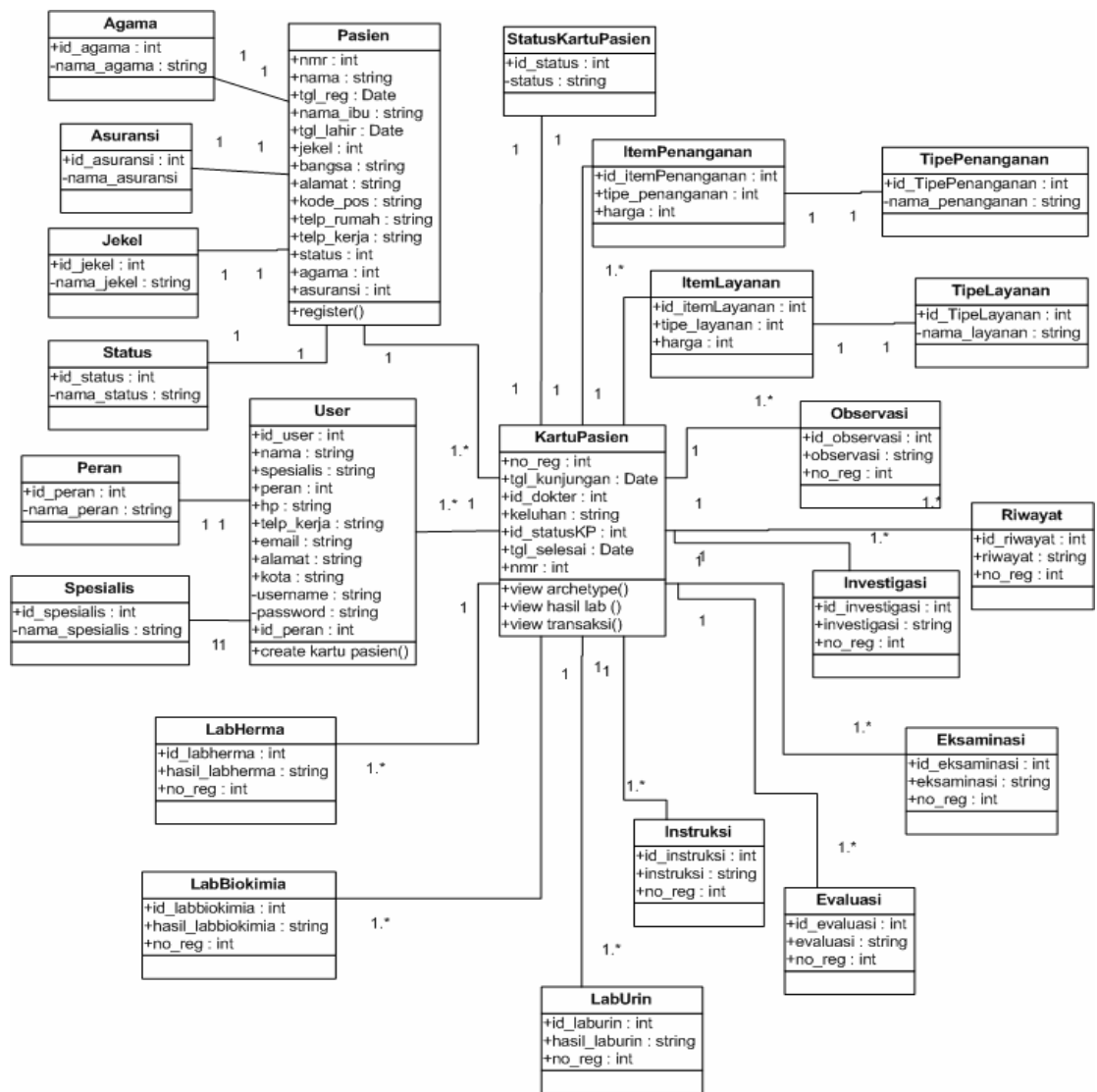

Figure 3: Class Diagram of Electronic Medical Record (EMR) Website System

Patient card is a representation of the patient's medical record. Patient card contains data on physician examination, laboratory results, and transaction. Cards patients had one table reference that is the status of the patient chart. Card status of these patients varies according to the condition the patient chart. One card patients have one patient card status. Doctor's examination can be written in the observation, history, investigation, examination, evaluation, and instruction. One card patient could have more than one doctor's examination. Laboratory results can be written in urine Lab, Biochemistry Lab, and hematology Lab, according to the laboratory where the patient examination. One card patient could have multiple laboratory results. Handling (action) item and services item is a table that represents the transaction. Handling item is for the cost if the patient gains examinations from a doctor, while the service item is the cost of the patient examination in the laboratory. One card patient has multiple transactions. This transaction has a reference to the type of treatment for handling items and service type for service items. One handling item has one type of treatment, as well as service item has one type of service.

### 3.3 Ontograf

*Ontograf* is ontology modeling that is used to describe the conception of an integrated knowledge base [2,3]. Ontograf is used to describe the ontology of the patient. The Ontograf of patient is shown in Fig.4.

From that figure it can be seen that the derivative of the Thing is the class of patients. This Class shows the table. Patients have 15 sub classes. Sub-class shows "possession" of the patient. The patient is a domain. Patients have four references: religion, sex, insurance, and status. Patients have a choice between religions, namely Buddhism, Catholicism, Hinduism, Islam, Protestant Christians and others. Patients have sex among women or men. Health Insurance patients have such insurance or Social Security. Patients also have a status among the divorced, married or single.

Patients have the status of a patient's card is a reference of the patient chart. There are four types of card status patients: wait-status, examined by a doctor, examined by a laboratory assistant, and complete. Patients have the type of service that is a reference to the handling item. This item includes the type of treatment such as dental treatment, eye doctors, and general practitioners. Patients have the type of service that is a reference to the services item such as service of hematology laboratory, urinalysis laboratory, and biochemistry laboratory.

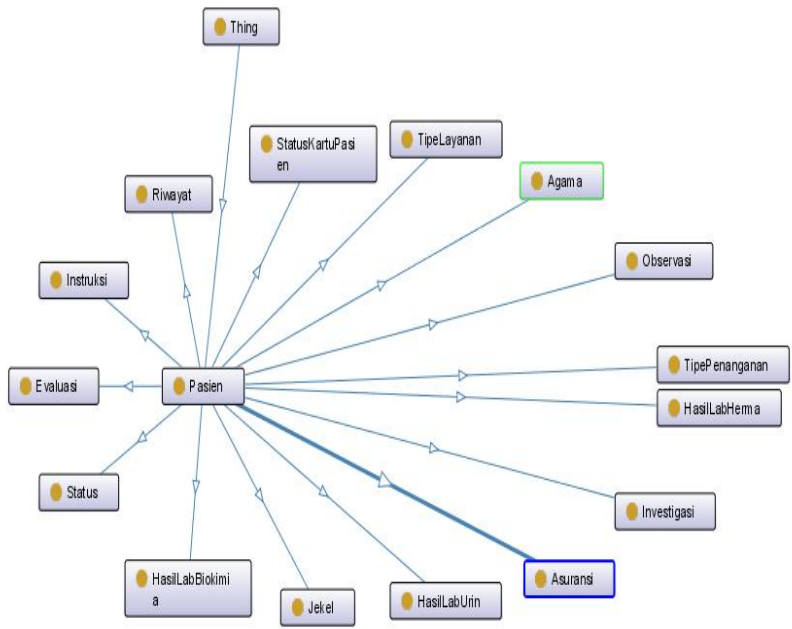

Figure 4: Ontograf of Patient

The patient has a doctor's examination including observation, evaluation, and instruction. Observations have four parts: observation, history, examination and investigation. Observations are observations that generate data that can be measured by numbers. Observation is the results directly to the patient. Evaluation is an observation that can not be measured as diagnostics. Evaluation performed by investigators who analyzed the observations results based on evidence and personal knowledge. Evaluation contains assessments, opinions and goals. Instruction is what to do by the doctor to the patient in the future. Instruction is something that should be assigned to agent investigator to give action to the patient.

**3.4 Navigation Structure**
The navigation structure is used to describe the flow of the application program. The used navigation structure is a mixture navigation structure, which is a combination of navigation structures linear, non-linear and hierarchical. Website navigation structure on the electronic medical record can be seen in Fig. 5.

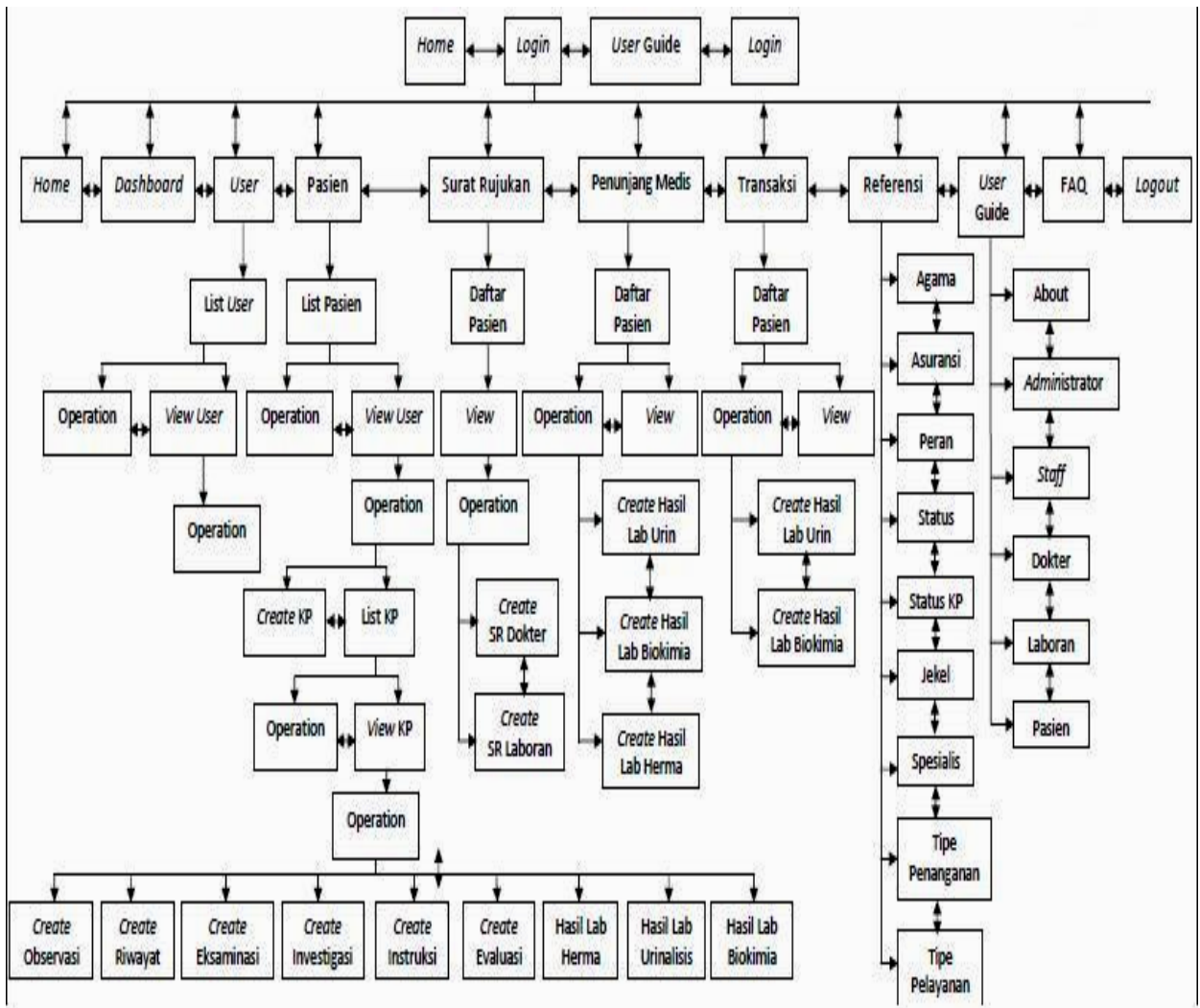

Figure 5: Navigation Structure of EMR Website System

From Fig. 5 it is shown when users access the website, there will be four menus: *home, login, user guide and FAQ*. Users can only access the Website after logging facility. Users who can login will see the home menu, dashboard, users, patients, referral letters, medical support, transactions, reference, user guide, FAQ and logout. From Fig.5 it is also shown that a user can move from one menu to another menu. Users who are logged in can view all the menus but not all users can use all the menus. Users can only use the menu in accordance with the role.

All users can access the home menu. Dashboard menu will only appear according to user role. Dashboard for each user is different from each other. Dashboard serves to facilitate the users in their activities in accordance with their respective roles. Menu user, patient, referral letters, medical support, and transactions can be accessed by all users except the patient, but users can only view and update, while to create, delete and set menus can only be done by an admin. When a user opens the menus, it will appear a list of users for user menu, and a list patients to other menus. When the user selects one user or patient, it will show the user or patient information and operation on the side menu bar. Operation contains create, update, delete, list, and manage.

When users access the patient's menu, it will display a list of patients. Menu patients can only be accessed by users other than the patient. Patients can only access the dashboard patients only. To access the patient chart, the user should select patients in the patient list. After performing patient demographics, patient list can be selected in the operation which is located in the side bar. One patient had more than one patient card, the card can be selected which patients will be seen. When viewing a patient's card, we can do operations such as creating observation, history, examination, investigation, instruction, evaluation, and viewing results of: hematology lab., urinalysis lab, and biochemistry lab. Patient card can only be made by the staff, and the card should only be filled out by a doctor.

The doctor can make a referral via the referral menu or a from a doctor dashboard. A doctor only can make a referral. The menu of medical support can be utilized by the laboratory assistant. On the menu of medical support the laboratory assistant can make the results of the laboratory where the laboratory assistant works. The laboratory assistant can also make laboratory results through the laboratory assistant dashboard. Transaction menu is a facility that can be used by staff. On the transaction menu staff can load transactions of patients according to the services or treatments that are gained by the patient. Staff can also make transactions through a dashboard staff.

For reference menu, user guide, and FAQs can be accessed by all users. Reference menu contains references to various categories that can be viewed by all users, but it can only be created, modified, deleted, and regulated by the admin. The menu contains a list of references id from a variety of categories, such as religion, insurance, role, status, and status of the patient chart, the type of service, type of treatment, sex, and specialists. For the user guide menu simply displays a list of general manual of the Website that is tailored to the user's role, the administrator, staff, physicians, laboratory, and patient. FAQ menu is the menu that contains frequently asked questions, in order to help answer user questions. Logout menu is used to exit from the website and return to the start menu.

## 4. Conclusion and Future Work

Web-based electronic medical record based on OpenEHR has been completed. Of the trials that have been conducted all the facilities contained in this website has been running well. The advantages of this application are to provide facilities to facilitate the processing of medical records by the actors that play a role in the processing of medical records such as staff, clinicians, and laboratory. This website also provides the facility that the patient can see his medical records and demographics, so that the patient can observe the development of his health. This website also uses OpenEHR's data standardization which includes information and service model for electronic medical records, demographics, and the archetype which allows software developers taking the logical structure as a universal functional interface, so it can facilitate the process of information by the recipient. It is because of using the interface with appropriate-purposed data presentation and data on computer screen of the same users. This website is a prototype of the system that is a complete electronic medical record system given the actual hospital. The website is still not developed by units that are in the hospital such as inpatient, outpatient, emergency room (ER), and others. Overall archetypes are not yet rated by the unit and are classified in general.

The website still has weaknesses. To overcome the drawbacks previously mentioned website, there should be further development to suit the needs of the units in the hospital. For the future the development of archetype can be customized with a template that exists in each unit in the hospital.